\documentclass[letterpaper,11pt]{article}

\usepackage[utf8]{inputenc}
\usepackage[T1]{fontenc}
\usepackage{cfr-lm}

\usepackage{relsize}
\usepackage{exscale}

\usepackage{bbold}

\usepackage{stmaryrd}
\usepackage{cite}
\usepackage[dvips]{graphicx}
\usepackage{graphpap}
\usepackage{ifthen}
\usepackage{enumerate}
\usepackage{amssymb}
\usepackage{mathrsfs}
\usepackage[errorshow]{tracefnt}
\usepackage{fancybox}
\usepackage{amsfonts}
\usepackage{amsmath}
\usepackage{amssymb}
\usepackage{multirow}
\usepackage{array}
\usepackage{mathtools}
\usepackage{soul}
\usepackage{pict2e}
\usepackage{mathabx}
\DeclareMathAlphabet{\mathpzc}{OT1}{pzc}{m}{it}

\usepackage{lscape}
\usepackage{xypic}
\usepackage{color}
\usepackage[
color,matrix,arrow]{xy}
\usepackage{setspace}

\usepackage{enumitem}

\usepackage{graphicx}

\makeatletter
\newcommand{\thickhline}{%
    \noalign {\ifnum 0=`}\fi \hrule height 1pt
    \futurelet \reserved@a \@xhline
}
\newcolumntype{'}{@{\hskip\tabcolsep\vrule width 1pt\hskip\tabcolsep}}
\makeatother
\newcolumntype{"}{@{\hskip\tabcolsep\vrule width 1.5pt\hskip\tabcolsep}}
\makeatother

\newcommand{\scr}{\mathscr}

\def\ie{{\it i.e.}}

\def\eg{{\it e.g.}}

\def\small#1{{\hbox{$#1$}}}
\def\fraction#1{\small{1\over#1}}
\def\fr{\fraction}
\def\Fraction#1#2{\small{#1\over#2}}
\def\Fr{\Fraction}

\def\boxit#1{\vbox{\hrule\hbox{\vrule\kern3pt
             \vbox{\kern3pt#1\kern3pt}\kern3pt\vrule}\hrule}}

\newcommand{\beq}{\begin{equation}}
\newcommand{\beqn}{\begin{equation*}}
\newcommand{\eeq}{\end{equation}}
\newcommand{\eeqn}{\end{equation*}}
\newcommand{\beqa}{\begin{eqnarray}}
\newcommand{\beqan}{\begin{eqnarray*}}
\newcommand{\eeqa}{\end{eqnarray}}
\newcommand{\eeqan}{\end{eqnarray*}}
\newcommand{\bdm}{\begin{displaymath}}
\newcommand{\edm}{\end{displaymath}}

\newcommand{\ba}{\begin{array}}
\newcommand{\ea}{\end{array}}

\newcommand\nn{\nonumber}

\newcommand\benu{\begin{enumerate}}
\newcommand\eenu{\end{enumerate}}
\newcommand\bit{\begin{itemize}}
\newcommand\eit{\end{itemize}}

\def\der'{\mathfrak{der}'\,}
\def\der{\mathfrak{der}\,}
\def\str'{\mathfrak{str}'\,}
\def\str{\mathfrak{str}\,}

\def\so{\mathfrak{so}}

\def\gl{\mathfrak{gl}}

\def\id{\mathbb{1}}   

\def\fg{{\mathfrak g}}
\def\fk{{\mathfrak k}}

\def\sh{\sharp}
\def\fl{\flat}

\def\*{\partial}

\def\tk{\widetilde k}

\def\RR{{\mathbb R}}

\def\LL{{\scr L}}

\def\compl{{\scriptscriptstyle\complement}}

\def\rep#1{\bf{#1}}

\RequirePackage{hyperref}

\numberwithin{equation}{section}

\begin{document}


\frenchspacing

\null\vspace{-28mm}

\includegraphics[height=2cm]{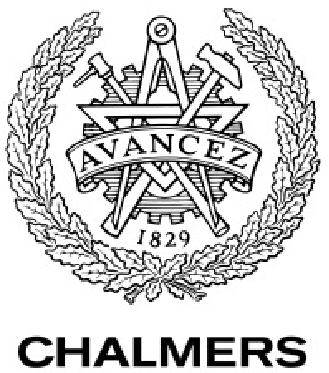}
\hspace{2mm}
\includegraphics[height=1.85cm]{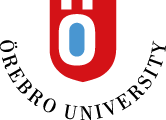}

\vspace{-12mm}
{\flushright Gothenburg preprint \\ 
}

\vspace{4mm}

\hrule

\vspace{16mm}


\thispagestyle{empty}

\begin{center}
  {\Large \bf \sc Teleparallelism in the algebraic approach}
  \\[3mm]
  {\Large \bf \sc to extended geometry}
    \\[10mm]
    
{\large
Martin Cederwall${}^1$ and Jakob Palmkvist${}^{2}$}

\vspace{10mm}
       {\footnotesize ${}^1${\it Department of Physics,
         Chalmers Univ. of Technology,\\
 SE-412 96 Gothenburg, Sweden}}

\vspace{2mm}
       {\footnotesize ${}^2${\it Department of Mathematics,
         \"Orebro Univ.,\\
 SE-701 82 \"Orebro, Sweden}}

\end{center}

\vfill

\begin{quote}
  
\textbf{Abstract:} 
Extended geometry is based on an underlying tensor hierarchy algebra.
We extend the previously considered $L_\infty$ structure of the local
symmetries (the diffeomorphisms and their reducibility)
to incorporate physical fields, field strengths and Bianchi
identities, and identify these
as elements of the tensor hierarchy algebra. The field strengths arise
as generalised torsion, so the naturally occurring complex in the
$L_\infty$ algebra is\\
$\ldots\leftarrow$ torsion BI's
$\leftarrow$ torsion $\leftarrow$ vielbein $\leftarrow$
diffeomorphism parameters $\leftarrow\ldots$\\
In order to obtain equations of motion, which are not in this complex,
(pseudo-)actions, quadratic in torsion, are given for a large class of
models. 
This requires considering the dual complex.
We show how local invariance under the compact subgroup locally
defined by a generalised metric arises as a ``dual gauge symmetry''
associated with a certain torsion Bianchi identity, generalising
Lorentz invariance in the teleparallel formulation of gravity. 
The analysis is performed
for a large class of finite-dimensional structure groups, with $E_5$
as a detailed example. The
continuation to infinite-dimensional cases is discussed.
\end{quote} 

\vfill

\hrule

\noindent{\tiny email:
  martin.cederwall@chalmers.se, jakob.palmkvist@oru.se}

\newpage

\tableofcontents

\section{Introduction}

Extended geometry \cite{Cederwall:2017fjm,Cederwall:2018aab,Cederwall:2019qnw,Cederwall:2019bai}, containing exceptional geometry \cite{Hull:2007zu,Pacheco:2008ps,Hillmann:2009pp,Berman:2010is,Berman:2011pe,Coimbra:2011ky,Coimbra:2012af,Berman:2012vc,Park:2013gaj,Cederwall:2013naa,Cederwall:2013oaa,Aldazabal:2013mya,Hohm:2013pua,Blair:2013gqa,Hohm:2013vpa,Hohm:2013uia,Hohm:2014fxa,Cederwall:2015ica,Butter:2018bkl,Bossard:2017aae,Bossard:2018utw,Bossard:2019ksx,Bossard:2021jix,Bossard:2021ebg} and double geometry \cite{Tseytlin:1990va,Siegel:1993xq,Siegel:1993bj,Hitchin:2010qz,Hull:2004in,Hull:2006va,Hull:2009mi,Hohm:2010jy,Hohm:2010pp,Jeon:2012hp,Park:2013mpa,Berman:2014jba,Cederwall:2014kxa,Cederwall:2014opa,Cederwall:2016ukd} as special examples, provides a means to promote  ``accidental'' duality symmetries to integral ingredients in a fundamental formulation of models containing gravity. Such symmetries include the Ehlers \cite{Ehlers:1957zz} and Geroch \cite{Geroch:1972yt} symmetries, and even the over-extended Belinskii–Khalatnikov–Lifshitz group \cite{BKL} close to a space-like singularity, so extended geometry carries a potential to reveal hidden information also about pure gravity.

It has become increasingly clear \cite{Palmkvist:2013vya,Palmkvist:2015dea,Cederwall:2019bai} that the proper framework for extended geometry is the tensor hierarchy algebras \cite{Palmkvist:2013vya,Carbone:2018xqq,Cederwall:2019qnw,Cederwall:2021ymp}. These Lie superalgebras give all information about gauge transformations, fields, field strengths etc., all in the framework of an $L_\infty$ algebra \cite{Cederwall:2018aab,Cederwall:2019bai}.
The purpose of the present paper is to report on progress along these lines. The field strengths naturally occurring in  the $L_\infty$ algebra arise as (generalised) torsion. Concepts like Riemann tensors have turned out generically not to be well defined in extended geometry, and spin connections are not uniquely determined by demanding covariant constancy of a vielbein. The natural and well defined complex to work with (and predicted by the tensor hierarchy algebras) contains
\begin{align*}
\cdots\ \leftarrow
\text{ torsion BI's }
\leftarrow \text{ torsion } 
\leftarrow
\text{ vielbein } \leftarrow
\text{ diffeomorphism parameters }
\leftarrow\ \cdots
\end{align*}
Equations of motion are to be found in the dual complex.
This points towards a teleparallel formulation being the most natural one, from the point of view of the algebraic foundations. Nevertheless, the dynamics of extended geometry has traditionally been formulated as coset dynamics on $G/K$, where local invariance under the compact subgroup $K$ of the structure group $G$ is manifest, but (generalised) diffeomorphism invariance is not. 
We now propose the opposite. Since torsion appears in the tensor hierarchy algebras, and is covariant under diffeomorphisms, we instead search for a teleparallel formulation of the dynamics, where diffeomorphisms invariance is manifest, but local $K$ invariance is not. We will show, for finite-dimensional structure groups, how it arises as what we choose to call ``dual gauge symmetry'', due to the presence of torsion Bianchi identities of a particular form. This may become important and informative when we turn to infinite-dimensional structure groups, and the peculiar ``extra'' elements extending the group $G$. 

We begin by reviewing the standard teleparallel formulation of gravity 
\cite{Einstein:1926,DeAndrade:2000sf} in Section \ref{TeleparallelGravity}, with particular focus on how the torsion Bianchi identity guarantees local Lorentz invariance.
Section \ref{TeleExtendedGeometry} identifies a certain Bianchi identity when $G$ is finite-dimensional.
A concrete example is then given in Section \ref{E5ExtGeometry}, where the complete construction of teleparallel dynamics is given for $E_5$ extended geometry.
A (pseudo-)action for a large class of extended geometries is the constructed in Section \ref{GeneralDynamics}.
We conclude by summarising the results and outlining future research in Section \ref{Conclusions}.

\section{Teleparallel gravity and dynamics on $GL(d)/SO(d)$\label{TeleparallelGravity}}

This section is a recapitulation of well known facts. Nevertheless, in order to clarify the mechanism behind local Lorentz invariance, which will be essential for the generalisation to extended geometry, we review the basics of teleparallel gravity.


We work in a $d$-dimensional space or space-time.
The gravity field is locally defined by a vielbein or frame $1$-form
$e^a=dx^me_m{}^a$. Indices in coordinate basis (``curved indices'') are labelled
$m,n,\ldots$ while $a,b,\ldots$ are Lorentz or orthogonal (``flat'')
indices. The metric is $g_{mn}=e_m{}^ae_n{}^b\eta_{ab}$, where $\eta$
is the flat metric with appropriate signature.

An affine connection in the curved basis is denoted $\Gamma=dx^m\Gamma_m$,
taking values in $\gl(d)$, \ie, having components $\Gamma_{mn}{}^p$. A
spin connection in the flat basis is $\omega=dx^m\omega_m$, taking
values in $\so(d)$ (with some signature), with components $\omega_{ma}{}^b$.
The compatibility relation for the vielbein and connections is
\begin{equation}
  D_me_n{}^a\equiv\*_me_n{}^a+\Gamma_{mn}{}^pe_p{}^a-\omega_{mb}{}^ae_n{}^b
  =0\;.
  \label{Compatibility}
\end{equation}
If the torsion part of the affine connection,
$\Theta_{mn}{}^p=2\Gamma_{[mn]}{}^p$ is fixed, the compatibility equation
determines the spin connection.

We will be particularly interested in the Weitzenb\"ock connection,
which is (minus) the right-invariant Maurer--Cartan $1$-form for the vielbein,
seen as a group element in $GL(d)$. From now on, we denote this
connection $\Gamma$, so
$\Gamma=-dee^{-1}$, or, in components, $\Gamma_{mn}{}^p=-\partial_me_n{}^ae_a{}^p$. This connection is flat --- the vanishing of its
curvature is expressed as the Maurer--Cartan equations
$d\Gamma+\Gamma\wedge\Gamma=0$.
Using the Weitzenb\"ock connection implies that the spin connection
vanishes through eq. \eqref{Compatibility}, which provides the basis
of the teleparallel formalism.
On the other hand, the Weitzenb\"ock connection carries a non-zero
torsion
$\Theta_{mn}{}^p=-2(\*_{[m}ee^{-1})_{n]}{}^p$, also known as the anholonomy
coefficients.

Starting from the Weitzenb\"ock connection, there are two ways of
formulating dynamics giving an action which is quadratic in the
derivative of the vielbein.
The principle employed is the invariance under both general coordinate
transformations and local orthogonal rotations.
Both can not be manifest at the same time (they are of course in the
Einstein--Hilbert action, which is not quadratic in first derivative
entities).

If one chooses to manifest general coordinate invariance,
one forms a Lagrangian from the torsion $\Theta$, which transforms as a
tensor (and, of course, the metric $g_{mn}$). Then, local rotations
are not manifest.
When the vielbein transforms as $\delta_\Lambda
e_m{}^a=-e_m{}^b\Lambda_b{}^a$ with $\Lambda\in\so(d)$,
the Weizenb\"ock connection transforms as
$\delta_\Lambda\Gamma_{mn}{}^p=(e\*_m\Lambda e^{-1})_n{}^p$.
Using the covariant constancy of the vielbein (\ref{Compatibility}), this can be written
$\delta_\Lambda\Gamma_{mn}{}^p=D_m\Lambda_n{}^p$, where indices on
$\Lambda$ have
been converted using the vielbein and its inverse. Thus,
\begin{align}
  \delta_\Lambda \Theta_{mn}{}^p=2D_{[m}\Lambda_{n]}{}^p\;.
  \label{DeltaLambdaT}
\end{align}
  As we will see
below, demanding invariance under local rotations, modulo a total
derivative, of a Lagrangian 
quadratic in torsion completely dictates its form. This leads to the
teleparallel formulation of gravitational dynamics.

If one, on the other hand, wants to manifest invariance under local
rotations, one notes that the transformation of the Weitzenb\"ock
connection $\Gamma_m$ only affects its component $Q_m$ in $\so(d)$ (as defined locally
by the involution induced by the metric). Defining
$\Gamma_m=P_m+Q_m$, where 
$P_m$ is a symmetric matrix, 
\ie, 
$P_{mn}{}^{p}=\Gamma_{m(np)}$ and $Q_{mn}{}^{p}=\Gamma_{m[np]}$, the coset
part $P_{mn}{}^p$ is invariant under local rotations.
On the other hand, it transforms inhomogeneously under general
coordinate transformations. If we define the inhomogeneous part of a
transformation with parameter $\xi$ as
$\Delta_\xi=\delta_\xi-L_\xi$, we have
$\Delta_\xi P_{mnp}=\*_m\*_{(n}\xi^qg_{p)q}$.
Demanding that a Lagrangian quadratic in $P$ is invariant under
general coordinate transformations, up to a total derivative,
determines its form. We refer to this formulation as the ``coset
model''.
A generalisation of this formulation has traditionally been used in
extended geometry.

Both the teleparallel formulation and the coset model are equivalent
to the Einstein--Hilbert action, and the actions differ from it and
from each other by total derivatives. The total derivative term
relating the teleparallel and coset formulations
will be invariant
neither under diffeomorphisms nor under local rotations.
We will determine the total derivative term below, and verify that
the general form of the action in extended geometry reduces to the one
for gravity.

It is straightforward to verify that in order for a $\Theta^2$ action to be
invariant under the transformation \eqref{DeltaLambdaT},
one is uniquely (up to an overall factor) led to the Lagrangian
\begin{align}
  \LL^{(\Theta)}=|e|\bigl(\frac18\Theta_{mnp}\Theta^{mnp}
  +\frac14\Theta_{mnp}\Theta^{mpn}-
  \frac12\Theta_{mn}{}^n\Theta^m{}_p{}^p
  \bigr)\;.
  \label{GravityTeleparallelAction}
\end{align}
One then gets
\begin{align}
  \delta_\Lambda\LL^{(\Theta)}
  =|e|\bigl(-\frac12\Theta_{mn}{}^pD_p\Lambda^{mn}
  +\Theta_{mp}{}^pD_n\Lambda^{mn}\bigr)
  =-\frac32|e|\Theta_{[mn}{}^pD_{p]}\Lambda^{mn}\;.
\end{align}
In order to show that this is a total derivative, one uses the Bianchi
identity for $\Theta$,
\begin{align}
  D_{[m}\Theta_{np]}{}^q=-\Theta_{[mn}{}^r\Theta_{p]r}{}^q\;,
\end{align}
together with the
observation that
\begin{align}
  \*_m(|e|v^m)&=|e|\bigl(\*_mv^m+(\*_mee^{-1})_n{}^nv^m\bigr)
  =|e|\bigl(\*_mv^m-\Gamma_{nm}{}^mv^n\bigr)\nn\\
  &=|e|\bigl(\*_mv^m-\Gamma_{mn}{}^mv^n-\Theta_{nm}{}^mv^n\bigr)
  =|e|\bigl(D_mv^m-\Theta_{mn}{}^nv^m\bigr)\;.
\end{align}
The $\Theta^2$ terms from the Bianchi identity and the partial integration
cancel, and we have
$\delta_\Lambda\LL^{(\Theta)}
=-\frac32\*_p\bigl(|e|\Theta_{mn}{}^{[p}\Lambda^{mn]}\bigr)$.
Using $[D_m,D_n]=\Theta_{mn}{}^pD_p$, this can be further rewritten as
\begin{align}
\delta_\Lambda\LL^{(\Theta)}=-\*_m\bigl(|e|D_n\Lambda^{mn}\bigr)\;.
\end{align}
Note that only the part of the Bianchi identity which is an antisymmetric tensor in lower indices is needed.
It reads
\begin{align}
D_p\Theta_{mn}{}^p+2D_{[m}\Theta_{n]p}{}^p=\Theta_{mn}{}^p\Theta_{pq}{}^q\;.
\end{align}

Now, it is obvious that this transformation can be cancelled by adding
the total derivative term $\*_m(|e|Q_n{}^{mn})$ to the Lagrangian.
This will however break the manifest invariance under diffeomorphisms,
since $Q$ is not a tensor. The resulting Lagrangian,
$\LL^{(P)}=\LL^{(\Theta)}+\*_m(|e|Q_n{}^{mn})$, will be invariant under
local $\so(d)$ rotations, but only invariant under diffeomorphisms
modulo a total derivative. It is straightforward to evaluate the
derivative in the extra term using the Maurer--Cartan equations
and check by explicit
calculation that $\LL^{(P)}$ contains only $P^2$ terms, and that terms
with $PQ$ and $Q^2$ vanish, as expected. The concrete expression for
$\LL^{(P)}$ is
\begin{align}
  \LL^{(P)}=\frac12P_{mnp}P^{mnp}-P_{mnp}P^{nmp}+P_{mn}{}^nP_p{}^{pm}
  -\frac12P_{mn}{}^nP^m{}_p{}^p\;.
\end{align}

\subsection{The coset model}

Before generalising this to extended geometry, we would like to check
that $\LL^{(P)}$ indeed coincides with the extended geometry
Lagrangian when specialised to gravity.
The general extended geometry Lagrangian, in models where ancillary
transformations are absent, is given by
\begin{align}
  \LL^{(\Pi)}&=\frac12G^{MN}\eta^{\alpha\beta}\Pi_{M\alpha}\Pi_{N\beta}
  -G^{PQ}t^\alpha{}_P{}^Mt^\beta{}_Q{}^N\Pi_{N\alpha}\Pi_{M\beta}\nn\\
  &-2(G^{-1}t^\alpha)^{MN}\Pi_{M\alpha}\pi_N
  -\frac{(\lambda,\lambda)}{(\lambda,\lambda)-\frac12}G^{MN}\pi_M\pi_N
  \label{LLPi}
\end{align}
Here, the generalised metric $G_{MN}$ is chosen to have the scaling
weight $-2((\lambda,\lambda)-\frac12)$ (instead of
$-2((\lambda,\lambda)-1)$ which is appropriate for a tensor), which
explains the absence of a determinant factor.
We also have
$-\frac12(\*_MGG^{-1})_N{}^P=\Pi_{M\alpha}t^\alpha{}_N{}^P+\pi_M\delta_N^P$.
Indices $\alpha,\beta,\ldots$ are adjoint, and $M,N,\ldots$ are
coordinate indices, corresponding to the highest/lowest weight module $R(\pm\lambda)$.
The matrices $t^\alpha{}_M{}^N$ are representation matrices for
$R(\lambda)$, and they are normalised so that
$\frac12\eta_{\alpha\beta}t^\alpha t^\beta=C_2(R(\lambda))\id
=\frac12(\lambda,\lambda+2\varrho)\id$.

Gravity in $d=n+1$ dimensions
is extended geometry with $\gl(n+1)\simeq
A_n\oplus\RR$ as structure algebra and the fundamental as coordinate representation.
The adjoint index $\alpha$ can be replaces by a (traceless) pair ${}_m{}^n$
The representation matrices are
\begin{align}
t_m{}^n{}_p{}^q=\delta_m^q\delta_n^p-\frac1{n+1}\delta_m^n\delta_p^q\;.
\end{align}
We have $(\lambda,\lambda)=\frac n{n+1}$. We rescale according to
$G_{mn}=|g|^{-\frac12}g_{mn}$. We can then relate $\Pi$ and $\pi$ to
$P$ as $\Pi_{mn}{}^p=P_{mn}{}^p-\frac1{n+1}\delta_n^pP_{mq}{}^q$ and
$\pi_m=-\frac{n-1}{2(n+1)}P_{mn}{}^n$.
Inserting these relations in $\LL^{(\Pi)}$ of eq. \eqref{LLPi} results
in an elimination of all $n$-dependence in the coefficients, and we
obtain exactly $\LL^{(\Pi)}=\LL^{(P)}$.

Forming the tensor $\varphi$ \cite{Cederwall:2019qnw}, which always is a sum of
projections on the ``big'' torsion modules (torsion not in
$R(\lambda)$).
We have the identity
\begin{align}
  \varphi^\alpha{}_{M,\beta}{}^N=\delta^\alpha_\beta\delta_M^N
  +f^\alpha{}_{\beta\gamma}t^\gamma{}_M{}^N
  -\frac1{(\lambda,\lambda)}(t^\alpha t_\beta)_M{}^N+\ell_{M\beta}{}^{N\alpha}\;.
\end{align}
Here, $\ell=0$, and insertion of the structure constants above yields
\begin{align}
  \varphi_{p}{}^q{}_{m,r}{}^{sn}
  =2\bigl(\delta_{mp}^{ns}\delta^q_r
  +\frac2n\delta_{[m}^q\delta_{p]}^{[n}\delta_{\mathstrut
      r}^{\mathstrut s]}\bigr)\;,
\end{align}
which obviously projects on the traceless torsion module.
Denoting the traceless part of torsion $\tilde\Theta_{mn}{}^p$, we have
$\tilde\Theta_{mn}{}^p=\varphi_n{}^p{}_{m,r}{}^{sq}\Gamma_{qs}{}^r$.

\subsection{Local Lorentz symmetry as a dual gauge symmetry}

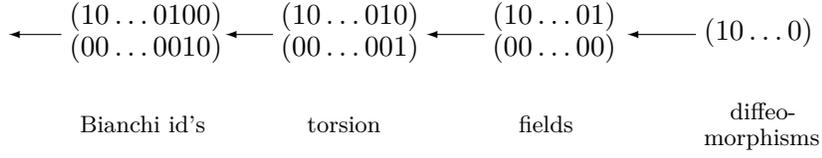
\begin{figure}
\begin{center}
\begin{picture}(330,70)(-20,-40)
\put(0,6){$(10\ldots0100)$}\put(0,-6){$(00\ldots0010)$}
\put(80,6){$(10\ldots010)$}\put(80,-6){$(00\ldots001)$}
\put(160,6){$(10\ldots01)$}\put(160,-6){$(00\ldots00)$}
\put(240,0){$(10\ldots0)$}
\put(-3,2){\vector(-1,0){20}}
\put(77,2){\vector(-1,0){18}}
\put(157,2){\vector(-1,0){22}}
\put(237,2){\vector(-1,0){26}}
\put(4,-35){\footnotesize Bianchi id's}
\put(90,-35){\footnotesize torsion}
\put(170,-35){\footnotesize fields}
\put(250,-30){\footnotesize diffeo-}
\put(240,-40){\footnotesize morphisms}
\end{picture}
\caption{The $W(d)$ complex of gravity.}
\label{WdFigure}
\end{center}
\end{figure}

The complex consisting of diffeomorphism parameters, (linearised) vielbein, torsion, torsion Bianchi identities, etc., can be identified with elements in the (finite-dimensional) Cartan-type superalgebra 
$W(d)=W(A_{d-1})$.  The $1$-bracket is depicted by the arrows in Figure
\ref{WdFigure}. There are no ancillary elements, which makes the embedding into $W(d+1)$ somewhat superfluous. If one wants to express the derivation as an algebraic operation it is however needed.

The procedure of formulating gravitational dynamics in the teleparallel formalism can be described as follows. First, the $1$-bracket is derived for the full non-linear fields. This gives the covariant field strengths, the torsion of the Weizenb\"ock connection, in terms of the vielbeins (not only its linear fluctuations around the unit matrix). Then, just as is done for a $p$-form field strength, a map $\star$ from the complex $C$ to its dual $C^\star$ is found. The equations of motion are then obtained as $d{\star}\Theta=0$, where $\Theta$ is the torsion. The map $\star:\,C\rightarrow C^\star$ is however not unique. It maps all $GL(d)$ modules to their duals. There is a ``canonical'' choice, the local Chevalley involution, which uses the metric 
on all indices. This is however not the choice implemented by the variation of the Lagrangian
\eqref{GravityTeleparallelAction}. Whenever there is a combination of upper and lower indices (or, equivalently, some tensor product with the adjoint), there is a choice of some linear combination of
$V_m{}^n\mapsto V_m{}^n$ and $V_m{}^n\mapsto g_{mp}g^{nq}V_q{}^p$.
For a specific choice of dualisation, the one implemented by the teleparallel Lagrangian,
the dual arrow from the dual of the antisymmetric Bianchi identity to the dual torsion becomes a gauge symmetry, \ie, the corresponding variation of the dual torsion in $C^\star$ is expressible as a transformation of the vielbein. Note that going to the dual complex is not associated to changing variables from the vielbein to some dual potential, but that the dual of the antisymmetric Bianchi identity is in the same position in the dual complex as a dual potential would be. We call this a dual gauge symmetry.

\section{Teleparallel extended geometry\label{TeleExtendedGeometry}}

We would like to generalise teleparallel formulation of gravity to extended geometry with structure group $G$. 
This will involve the identification of a specific antisymmetric
Bianchi identity, responsible for the invariance under local rotations
in the compact subgroup $K\subset G$.
In the teleparallel formulation, all physical entities will transform
covariantly under generalised diffeomorphisms, as described in Section
\ref{XGeoQuantities}.

\subsection{Modules of Bianchi identities in the tensor hierarchy algebra}

\begin{figure}
  \centerline{\includegraphics{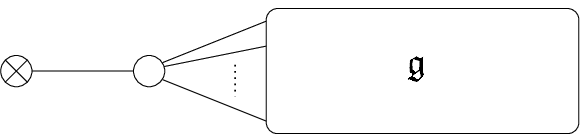}\hskip5mm\includegraphics{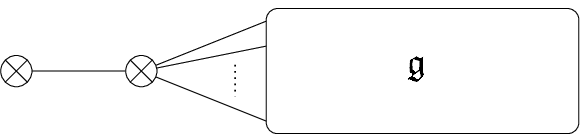}}
  \caption{\it Two equivalent Dynkin diagrams for ${\mathscr B}(\fg^+)$, $W(\fg^+)$
    and $S(\fg^+)$. Removing the ``grey'' node in the first diagram
    yields the Dynkin diagram of $\fg^+$.} 
  \label{DynkinFigure}
\end{figure}

\begin{table}
  \begin{align*}
  \xymatrix@=.4cm{
    \ar@{-}[]+<2.2em,1em>;[ddddd]+<2.2em,-1em>
    \ar@{-}[]+<-0.8cm,-1em>;[rrr]+<3cm,-1em>
   &\ar@{-}[]+<4.2em,1em>;[ddddd]+<4.2em,-1em> p=-2
    &\ar@{-}[]+<3.2em,1em>;[ddddd]+<3.2em,-1em> p=-1
    & \ar@{-}[]+<2.5em,1em>;[ddddd]+<2.5em,-1em> p=0 
    &p=1\\
q=2&&&&L_{\alpha M}^\sh\\
q=1&\ldots&\Phi^\sh_{\alpha}{}^{M}&T_\alpha^\sh& E_M^\sh\quad L_{\alpha M}
       \\ 
q=0&A^{MN}\,S^{MN}\,S'^{MN}&\Phi_{\alpha}{}^M\quad F^M
&\tk\quad{T_\alpha}
           & E_M\\ 
q=-1 &\ldots& F^{\fl M}& {e_0} &\\
q=-2 &\ldots&&& \\
  }
\end{align*}
  \caption{\it Basis elements for $S(\fg^+)$, for finite-dimensional $\fg$, at $p=-2,-1,0,1$. At $p=-2$, the list is incomplete --- only the modules used in this paper are given.}
\label{GeneralSTableBasis}
\end{table}

The superalgebra used in the identification of local symmetries,
fields, field strengths etc. in extended geometry with structure
algebra $\fg$ is the tensor hierarchy algebra $S(\fg^+)$
\cite{Cederwall:2019qnw,Cederwall:2019bai}. This superalgebra is
obtained as a double extension of $\fg$: first by a bosonic node,
resulting in the Kac--Moody algebra $\fg^+$. The Dynkin diagram of
$\fg^+$ is obtained by connecting the extending node to the Dynkin
diagram of $\fg$ by lines corresponding to the Dynkin labels of
$\lambda$, where the coordinate module is the lowest weight module
$R(-\lambda)$.
Then, the Dynkin diagram is further extended by attaching a fermionic
(``grey'') node to the first node. One obtains the first diagram in
Figure \ref{DynkinFigure}. The second diagram describes the same
algebra, after a ``fermionic Weyl reflection''.
The same diagrams, and the same Cartan matrices, apply to $W(\fg^+)$ and the
contragredient Borcherds superalgebra ${\scr B}(\fg^+)$.
For the full construction, we refer to
refs. \cite{Carbone:2018xqq,Cederwall:2019qnw}.

We will sometimes refer to the double grading of $S(\fg^+)$ with
respect to its two leftmost nodes (in any of the two diagrams).
Let us use a numbering where these two nodes are number $0$ and $1$,
and the remaining ones $2,\ldots,r$, with $r-1=\hbox{rank}\,\fg$.
(This numbering is the same as in \cite{Carbone:2018xqq}, but different from the one in
\cite{Cederwall:2019qnw,Cederwall:2021ymp}.)
We then denote the degree with respect to nodes $0$ and $1$ in the
right diagram by $-q$ and $p$. Equivalently, the gradings with respect
to nodes $0$ and $1$ in the left diagram are $p-q$ and $p$.
Non-ancillary elements in the $L_\infty$ algebra of extended geometry
reside at $q=0$, while ancillary ones are found at $q=1$. Their ghost
numbers are $p+q$.

For finite-dimensional $\fg$, ancillary terms only appear at positive
ghost numbers. In order to examine the content of torsion, which is a
field strength, \ie, ghost number $-1$, the subalgebra at $q=0$
suffices. This is $W(\fg)$, described by the right diagram in Figure~\ref{DynkinFigure} with the leftmost node removed.
Its Chevalley generators contain the standard ones $h_a,e_a,f_a$,
for $a=0,\ldots, r$ (generating the subalgebra $\mathscr B(\fg)$), but also
fermionic 
generators $f_{0i}$ for $i=2,\ldots,r$. 
They satisfy $[e_0,f_{0i}]=h_i$, why it is natural to set $f_{00}=f_{0}$.
Furthermore, $[e_1,f_{0i}]=0$, but $[e_i,f_{0j}]\neq0$ if the nodes $i$ and $j$
are connected.
We refer to refs.
\cite{Carbone:2018xqq,Cederwall:2019qnw} for a complete list of the defining identities.

We will now show that the leading antisymmetric and the subleading
symmetric modules are present at level $-2$ in $W(\fg)$. 
The proof relies on the identity $[e_1,[e_i,f_{0j}]]=0$,\linebreak which is obvious if the nodes $i$ and $j$ are disconnected, and otherwise by noting 
that $[e_1,[e_i,f_{0j}]]$ is proportional to $2[e_1,[e_i,f_{0i}]]=[e_1,[e_i,[e_i,[f_i,f_{0i}]]]]$, which vanishes by the Serre relations.

Let $f_{0\lambda}$ be the linear combination of $f_{0i}$ generators,
where the coefficients are the same as when $\lambda$ is expressed as a linear combination of the corresponding simple roots
$\alpha_i$.
Consider the element $[f_{0\lambda},[f_1,f_{00}]]$ in $W(\fg)$. The adjoint action on it of $e_i$ ($i=2,\ldots,r$) gives
\begin{align}
[e_i,[f_{0\lambda},[f_1,f_{00}]]]&=[[e_i,f_{0\lambda}],[f_1,f_{00}]]+[f_{0\lambda},[e_i,[f_1,f_{00}]]]\\
&=[[[e_i,f_{0\lambda}],f_1],f_{00}]+[f_1,[[e_i,f_{0\lambda}],f_{00}]]\,,\nn\\
&=-[[e_i,[f_1,f_{0\lambda}]],f_{00}]+[f_1,[e_i,[f_{0\lambda},f_{00}]]]-[f_1,[f_{0\lambda},[e_i,f_{00}]]]\;,\nn
\end{align}
which is zero since $[e_i,[f_1,f_{0\lambda}]]=[f_{0\lambda},f_{00}]=[e_i,f_{00}]=0$. In the first step we have also used that
$[e_i,[f_1,f_{00}]]]=0$. The fact that $[e_i,[f_1,f_{0\lambda}]]=0$ follows directly if $(\alpha_i,\lambda)=0$, and
otherwise using $[f_1,[f_1,[e_1,[e_i,f_{0\lambda}]]]]=0$, which in turn follows from $[e_1,[e_i,f_{0\lambda}]]=0$.
Acting with $e_1$ on $[f_{0\lambda},[f_1,f_{00}]]$, we get
\begin{align}
[e_1,[f_{0\lambda},[f_1,f_{00}]]]=[f_{0\lambda},[e_1,[f_1,f_{00}]]]=[f_{0\lambda},[h_1,f_{00}]]=[f_{0\lambda},f_{00}]=0\;.
\end{align}
Finally, acting with $e_0$ on $[f_{0\lambda},[f_1,f_{00}]]$, we get
\begin{align}
[e_0,[f_{0\lambda},[f_1,f_{00}]]]&=[h_\lambda,[f_1,f_{00}]]-[f_{0\lambda},[e_0,[f_1,f_{00}]]]\nn\\
&=(\lambda,\lambda)[f_1,f_{00}]+[f_{0\lambda},f_1]\;,
\end{align}
which is obviously nonzero. Thus $[f_{0\lambda},[f_1,f_{00}]]$ is a nonzero element that is annihilated by $e_1,e_2,\ldots,e_r$,
and it has the correct $\fg$-weight.
A covariant construction of the highest weight module that it generates is given
by the leading antisymmetric module in
$t^\alpha{}_P{}^{[M}[F^P,\Phi_\alpha{}^{N]}]$, where $F$ and $\Phi$
are elements at level $-1$ in $W(\fg)$, forming a basis for the
torsion module (for a precise characterisation of this module, see the
following subsection). Also the subleading antisymmetric modules will
be present.
In a similar fashion, subleading symmetric generators are formed as
$[F^M,F^N]$ and as $t^\alpha{}_P{}^{(M}[F^P,\Phi_\alpha{}^{N)}]$. The
subleading property of the latter is due to the properties of the
``big'' torsion module (spanned by $\Phi$), \ie, of the tensor
$\varphi$ in the following subsection.

In general, there will be many more irreducible modules for torsion
Bianchi identities at level $-2$ in $W(\fg)$ (see \eg\ Figure
\ref{EFiveComplex} for the $E_5$ model). They will not be
relevant to the discussion of local symmetries.

\subsection{Vielbein, torsion and Bianchi identities\label{XGeoQuantities}}

The coset $G/K\times\RR$ is parametrised by a generalised vielbein
$E_M{}^A$. For convenience, it can be assigned a (non-zero) scaling weight $w$. The tensorial value of $w$ for a covector is $1-(\lambda,\lambda)$ \cite{Cederwall:2017fjm}.
The Weizenb\"ock  connection is defined as the right-invariant Maurer--Cartan form
\begin{align}
\Gamma_M=-\*_MEE^{-1}\;,
\end{align}
and it is decomposed as
\begin{align}
\Gamma_{MN}{}^P=t_{\alpha N}{}^P\Gamma_M{}^\alpha+w\delta_N^P\gamma_M\;.
\end{align}
The vielbein is covariantly constant, $D_ME_N{}^A=\*_ME_N{}^A+\Gamma_{MN}{}^PE_P{}^A=0$, without any spin connection.
Using the covariant transformation of $E$ under generalised
diffeomorphisms \cite{Cederwall:2017fjm},
\begin{align}
\delta_\xi E_M{}^A=\xi^N\*_NE_M{}^A+t^\alpha{}_M{}^Qt_{\alpha N}{}^P\*_P\xi^NE_Q{}^A
+w\*_N\xi^NE_M{}^A\;,
\end{align}
the inhomogeneous parts of the transformations of the connection components follow:
\begin{align}
\Delta_\xi\Gamma_M{}^\alpha&=-t^\alpha{}_N{}^P\*_M\*_P\xi^N\;,\nn\\
\Delta_\xi\gamma_M&=-\*_M\*_N\xi^N\;.
\end{align}
Note that, unlike other connections, the Weizenb\"ock connection
satisfies a section constraint on its first index.

In ref. \cite{Cederwall:2019qnw}, invariant tensors $\varphi$ and $\ell$ were introduced, that occur in the structure constants of the relevant tensor hierarchy algebra $S(\fg^+)$.
The tensor $\varphi$ projects (with some non-zero weights) on the modules occurring in the ``big'' torsion module, consisting of the irreducible modules with highest weights
$\lambda+\gamma$, where $\gamma$ belongs to the set of highest $\fg$ roots with $(\lambda,\gamma)=0,-2,-3,\ldots,-(\lambda,\theta)$. The tensor $\ell$, which is associated with the presence of ancillary transformation, projects on the subset of these with $(\lambda,\gamma)\leq-2$.
When ancillary transformations are not present, $\ell=0$ and the highest weights appearing in the big torsion module are only $\lambda+\gamma$ with $(\lambda,\gamma)=0$.
Much of the considerations below, in particular the action of Section \ref{GeneralDynamics} will concern 
this situation, and also, for simplicity, cases where the big torsion module is irreducible, \ie, when there is a single highest root with $(\lambda,\gamma)=0$.

We then immediately find the ``big'' and ``small'' torsion as
\begin{align}
\Theta_M{}^\alpha&=\varphi^\alpha{}_{M,\beta}{}^N\Gamma_N{}^\beta\;,\nn\\
\theta_M&=t_{\alpha M}{}^N\Gamma_N{}^\alpha-(\lambda,\lambda)\gamma_M\;.
\end{align}
This holds for finite-dimensional $\fg$. 


We denote projection on the leading symmetric module $R(2\lambda)$ by $\langle MN\rangle$, 
and on the leading antisymmetric module(s) $R(2\lambda-\alpha_i)$ by $\{MN\}$.

The tensors $\varphi$ and $\ell$ turn out to respect a ``remarkable identity'' \cite{Cederwall:2019qnw}, appearing as a Jacobi identity in the tensor hierarchy algebra $S(\fg^+)$:
\begin{align}
\varphi^\beta{}_{M,\alpha}{}^N-\ell_{\alpha M}{}^{\beta N}
=\delta_\alpha^\beta\delta_M^N-f_\alpha{}^{\beta\gamma}t_{\gamma M}{}^N
-\fr{(\lambda,\lambda)}(t^\beta t_\alpha)_M{}^N\;,
\end{align}
or, in a notation with fundamental indices suppressed:
\begin{align}
\varphi^\beta{}_\alpha-\ell_\alpha{}^\beta
=\delta_\alpha^\beta-f_\alpha{}^{\beta\gamma}t_\gamma
-\fr{(\lambda,\lambda)}t^\beta t_\alpha\;.\label{RemarkId}
\end{align}

One defining relation for the tensor $\varphi$ is that it respects the level $2$ Serre relations, which means that
\begin{align}
(\varphi^\alpha{}_\beta\otimes t^\beta)_{MN}{}^{\langle PQ\rangle}=0\;.
\label{tPhiZero}
\end{align}
Since $\ell$ projects on a subset of the same modules, also
$(\ell_\alpha{}^\beta\otimes t_\beta)_{MN}{}^{\langle PQ\rangle}=0$.
In addition we have
$(\ell_\alpha{}^\beta\otimes t_\beta)_{MN}{}^{\{PQ\}}=0$.
Inserting these identities in the remarkable identity \eqref{RemarkId} and using the section constraint then gives
\begin{align}
(t_\beta\otimes\varphi^\beta{}_\alpha)_{MN}{}^{\langle PQ\rangle}
&=\bigl(-(1-\fr{(\lambda,\lambda)})f_\alpha{}^{\beta\gamma}t_\beta\otimes t_\gamma
+t_\alpha\otimes1-1\otimes t_\alpha\bigr){}_{MN}{}^{\langle PQ\rangle}\;,\nn\\
(t^\beta\otimes\ell_\beta{}^\alpha)_{MN}{}^{\langle PQ\rangle}
&=\bigl(-f_\alpha{}^{\beta\gamma}t_\beta\otimes t_\gamma
-t_\alpha\otimes1+1\otimes t_\alpha\bigr){}_{MN}{}^{\langle PQ\rangle}\;,\label{PhiEllRels}\\
(t_\beta\otimes\varphi^\beta{}_\alpha)_{MN}{}^{\{PQ\}}
&=\bigl(-(1-\fr{(\lambda,\lambda)})f_\alpha{}^{\beta\gamma}t_\beta\otimes t_\gamma+t_\alpha\otimes1+\Fr{2-(\lambda,\lambda)}{(\lambda,\lambda)}1\otimes t_\alpha\bigr){}_{MN}{}^{\{PQ\}}\;,\nn
\end{align}
(The second of these equations governs the appearance of ancillary transformations in the commutator of two generalised diffeomorphisms.)
Define $\psi^\alpha{}_\beta=\varphi^\alpha{}_\beta-(1-\fr{(\lambda,\lambda)})\ell^\alpha{}_\beta$.
Then, combining the first two relations in eq. \eqref{PhiEllRels} gives
\begin{align}
(t_\beta\otimes\psi^\beta{}_\alpha)_{MN}{}^{\langle PQ\rangle}
=(2-\fr{(\lambda,\lambda)})(t_\alpha\otimes1-1\otimes t_\alpha)_{MN}{}^{\langle PQ\rangle}\;.
\label{psiId}
\end{align}
Taking the antisymmetric part of the third relations gives
\begin{align}
(t_\beta\otimes\varphi^\beta{}_\alpha)_{[MN]}{}^{\{PQ\}}
=\fr{(\lambda,\lambda)}(t_\alpha\otimes1+1\otimes t_\alpha)_{[MN]}{}^{\{PQ\}}\;.
\label{phiId}
\end{align}

These last equations will be useful for finding the relevant Bianchi identity.
There is always a torsion Bianchi identity in an antisymmetric module.
This can be seen as follows. 
Define the big torsion as a different linear combination of irreducible modules than the one above:
\begin{align}
\Psi_M{}^\alpha
=\Bigl(\varphi^\alpha{}_{M,\beta}{}^N
-\bigl(1-\fr{(\lambda,\lambda)}\bigr)\ell^\alpha{}_{M,\beta}{}^N\Bigr)\Gamma_N{}^\beta
=\psi^\alpha{}_{M,\beta}{}^N\Gamma_N{}^\beta
\;.\label{PsiTorsionDef}
\end{align}
We then form
\begin{align}
B_{MN}=t_{\alpha[M}{}^PD_{|P|}\Psi_{N]}{}^\alpha
+2\bigl(2-\fr{(\lambda,\lambda)}\bigr)D_{[M}\theta_{N]}\;,
\end{align}
\ie,
\begin{align}
B=\Fr{1-\sigma}{2}(t_\alpha\otimes1)D\otimes\Psi^\alpha
+2\bigl(2-\fr{(\lambda,\lambda)}\bigr)D\wedge\theta\;,
\end{align}
(Here, $\sigma$ is the permutation operator; ${1-\sigma\over2}$
projects on the antisymmetric part of a tensor product.)
Thanks to eq. \eqref{psiId}, the part without the connection terms is
\begin{align}
&\Fr{1-\sigma}{2}(t_\alpha\otimes1)\*\otimes\Psi^\alpha
+2\bigl(2-\fr{(\lambda,\lambda)}\bigr)\*\wedge\theta\nn\\
&\quad=\Fr{1-\sigma}{2}(t_\alpha\otimes\psi^\alpha{}_\beta)\*\wedge\Gamma^\beta
+\bigl(2-\fr{(\lambda,\lambda)}\bigr)(t_\alpha\otimes1+1\otimes t_\alpha)\*\wedge\Gamma^\alpha\;.
\end{align}
We express the right hand side using the Maurer--Cartan equation for $\Gamma$ and reinsert the connection terms in the covariant derivatives. The result is tensorial, and therefore expressible in terms of torsion.

A lengthy calculation, making use of eqs. \eqref{psiId} and \eqref{phiId} shows that
\begin{align}
B&=-(t_\alpha\otimes1)\theta\otimes\Psi^\alpha\nn\\
&\quad\,-(1-\fr{(\lambda,\lambda)})\Bigl((t_\alpha\otimes1+1\otimes t_\alpha)t_\delta\otimes\ell^\delta{}_\beta
-{1\over2}f_{\alpha\beta}{}^\gamma t_\delta\otimes\ell^\delta{}_\gamma\Bigr)
\Gamma^\alpha\wedge\Gamma^\beta\\
&\quad\,+(2-\fr{(\lambda,\lambda)})(1-\fr{(\lambda,\lambda)})(\lambda,\lambda)
(t_\beta\otimes\ell^\beta{}_\alpha)\,\gamma\wedge\Gamma^\alpha\;,\nn
\end{align}
where projection from the left on the antisymmetric part by ${1-\sigma\over2}$ is understood.
The first term, which is expressed in terms of torsion, is precisely the desired one.
The remainder, still expressed in terms of connection, should be expressible in terms of torsion.
We have not found a straightforward way of doing this in general.
However, these terms do not contribute to the leading antisymmetric module, thanks to
$(t_\beta\otimes\ell^\beta{}_\alpha)_{\{MN\}}{}^{PQ}=0$.
The Bianchi identity in the leading antisymmetric module therefore is
\begin{align}
(D_P+\theta_P)X_{MN}{}^P\;,\label{LeadingBI}
\end{align}
where
\begin{align}
X_{MN}{}^P=t_{\alpha\{M}{}^P\Theta_{N\}}{}^\alpha
+2(2-\fr{(\lambda,\lambda)})\delta_{\{M}^P\theta^{\mathstrut}_{N\}}\;.
\end{align}
(Restricting to the leading module also implies that the terms with $\ell$ in eq. \eqref{PsiTorsionDef} drop out, $t_{\alpha\{M}{}^P\Psi_{N\}}{}^\alpha=t_{\alpha\{M}{}^P\Theta_{N\}}{}^\alpha$.)

The form of eq. \eqref{LeadingBI} is precisely the one needed in order to implement local invariance. The covariant divergence of a vector with weight $(\lambda,\lambda)$ is 
\begin{align}
D_Mv^M=(\*_M-\theta_M)v^M\;,
\end{align}
so the Bianchi identity can be written $\*_P(K^{MN}X_{MN}{}^P)=0$, where 
$K$ is covariantly constant and carries weight $2(\lambda,\lambda)-1$.

If the antisymmetric module is irreducible, \ie, consists only of its leading part, this Bianchi identity is enough. If not, it is desirable to find subleading Bianchi identities with the same property. We have not been able to produce a general derivation of such identities. It is however reassuring that they are present \eg\ when the coordinate module is the adjoint (see Section \ref{Conclusions}).

\section{An example: $E_5$ geometry\label{E5ExtGeometry}}

As a preparation, and a proof of concept, we consider a finite-dimensional case with irreducible $\Theta$.
We choose $E_5\simeq D_5$, where the coordinate module is a chiral spinor.

\subsection{$E_5$ decomposition of $S(E_6)$}

\begin{figure}
\begin{center}
\begin{picture}(330,90)(-90,-40)
\put(-120,18){$(11000)$}\put(-120,6){$(00002)$}
\put(-120,-6){$(00100)$}\put(-120,-18){$2(10000)$}
\put(-60,6){$(10010)$}\put(-60,-6){$(00001)$}
\put(0,6){$(01000)$}\put(0,-6){$(00000)$}
\put(60,0){$(00010)$}
\put(120,0){$(10000)$}
\put(180,0){$(00001)$}
\put(180,40){$(00001)$}
\put(240,0){$(01000)$}
\put(240,40){$(01000)$}
\put(-122,2){\vector(-1,0){20}}
\put(-62,2){\vector(-1,0){20}}
\put(-2,2){\vector(-1,0){20}}
\put(58,2){\vector(-1,0){20}}
\put(118,2){\vector(-1,0){20}}
\put(178,2){\vector(-1,0){20}}
\put(238,42){\vector(-1,0){20}}
\put(238,2){\vector(-1,0){20}}
\put(298,2){\vector(-1,0){20}}
\put(298,42){\vector(-1,0){20}}
\put(198,35){\vector(0,-1){24}}
\put(258,35){\vector(0,-1){24}}
\put(-125,-35){\footnotesize Bianchi id's}
\put(-58,-35){\footnotesize torsion}
\put(5,-35){\footnotesize fields}
\put(65,-30){\footnotesize diffeo-}
\put(58,-40){\footnotesize morphisms}
\put(115,-35){\footnotesize reducibility}
\end{picture}
\caption{The complex of $E_5$ extended geometry}
\end{center}
\label{EFiveComplex}
\end{figure}
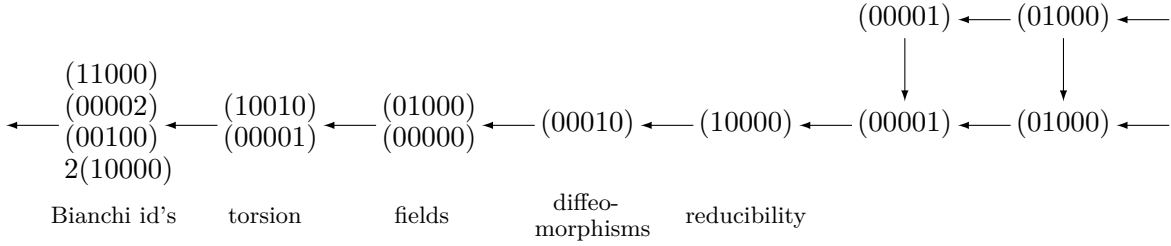

The fields at all ghost numbers are obtained as certain elements in an $E_5$-covariant bigrading of the tensor hierarchy algebra $S(E_6)$, as described in ref. \cite{Cederwall:2019bai}. The non-ancillary ones fill out the subalgebra $W(E_5)$, the lower line in Figure \ref{EFiveComplex}. Their ghost numbers coincide with the level of the grading of $W(E_5)$ with respect to its fermionic node. Ancillary ghosts, starting from ghost number $4$, are found in the upper line, and will not be considered here.
Importantly, among the torsion Bianchi identities at level $-2$, we find a $3$-form, \ie, an antisymmetric bispinor.

\subsection{$G=Spin(5,5)$ and $K=Spin(5)\times Spin(5)$}

We use different conventions for indices in this Section compared to the rest of the paper.
Spinor indices are $\alpha,\beta,\ldots$. Vector indices are $a,b,\ldots$.
An adjoint index is $[ab]$.
In order to have conventions that easily translate to the general framework (\eg, in terms of normalisation of generators, the Killing metric etc.) we choose representation matrices in the vector and spinor representations:
\begin{align}
(t_{ab})_c{}^d&=2\eta_{c[a}\delta_{b]}^d\;,\nn\\
(t_{ab})_\alpha{}^\beta&={1\over2}(\gamma_{ab})_\alpha{}^\beta \label{RepMatrEq}\;.
\end{align}
We also use the Killing metric ($A=[ab]$, $B=[cd]$)
\begin{align}
\eta^{AB}=-{1\over2}\eta^{[c[a}\eta^{b]d]}\;.
\end{align}
If we use the vector metric to raise and lower indices, this implies a factor $-{1\over2}$ in any contraction of adjoint indices. (This can be avoided, but at the price of having factors of $\sqrt2$ in the representation matrices.)

We can check the conventions in the section constraint, which only has a symmetric part,
$(\eta^{AB}t_A\otimes t_B-(\lambda,\lambda))(\partial\otimes\partial)=0$.
We then get ($\langle\alpha\beta\rangle$ denoting projection on $S_2^\compl$, the $5$-form)
\begin{align}
(t^A\otimes t_A)_{\alpha\beta}{}^{\langle\gamma\delta\rangle}
&=-{1\over8}(\gamma_{ab})_\alpha{}^{\langle\gamma}(\gamma_{ab})_\beta{}^{\delta\rangle}
=-{1\over8}(\gamma_a\gamma_b)_\alpha{}^{\langle\gamma}
           ((\gamma^a\gamma^b)_\beta{}^{\delta\rangle}-\eta^{ab}\delta_\beta^{\delta\rangle})\nn\\
&={5\over4}\delta_\alpha^{\langle\gamma}\delta_\beta^{\delta\rangle}\;,
\end{align}
and $(\lambda,\lambda)={5\over4}$ is the length squared of a spinor highest weight.
The structure constants are 
\begin{align}
f_{ab,cd}{}^{ef}=8\eta^{\mathstrut}_{[c[a}\delta_{b]}^{[e}\delta_{d]}^{f]}\;.
\end{align}

The group $G$ is broken to $K$ by the introduction of a generalised metric $G_{\alpha\beta}$. 
Since it is the $5$-form module that contains the $K$ singlet, we can write
$G_{\alpha\beta}={1\over 2\cdot5!}(\gamma^{abcde})_{\alpha\beta}G_{abcde}$. This simply means that the involution defined by the metric is in the same $G$-orbit as the Chevalley involution, the Cartan involution corresponding to the split real form. Also,
$(G^{-1})^{\alpha\beta}={1\over 2\cdot5!}(\gamma_{abcde})^{\alpha\beta}(G^{-1})^{abcde}$. 
We have ${1\over4!}G_{acdef}(G^{-1})^{bcdef}=\delta_a^b$.
A metric on the vector module (different from the $Spin(5,5)$-invariant one $\eta_{ab}$) is
$H_{ab}={1\over4!}G_a{}^{cdef}G_{bcdef}$. It lies in $(20000)$, \ie, $\eta^{ab}H_{ab}=0$.
Identities include 
$H_a{}^f(G^{-1})_{fbcde}=G_{abcde}$. Such relations can generically be found by excluding any module, in this case $(10000)\otimes(00011)$, not containing a $K$ singlet.

\subsection{Vielbeins, connection and torsion}

We parametrise the coset $(Spin(5,5)\times\RR)/(Spin(5)\times Spin(5))$ by a generalised vielbein
$E_\alpha{}^{\tilde\alpha}$, where $\alpha$ is a $Spin(5,5)$ chiral spinor index, and $\tilde\alpha$ 
a ``flat'' $Spin(5)\times Spin(5)$ bispinor index. 
Let $E$ transform under generalised diffeomorphisms with a weight $w$, the value of which we for now keep open. By the weight of a field $X$ in some $Spin(5,5)$ module we mean the coefficient $w_X$ in
\begin{align}
\delta_\xi X=\xi^\alpha\*_\alpha X+(t^A)_\alpha{}^\beta\*_\beta\xi^\alpha t_A\cdot X
+w_X\*_\alpha\xi^\alpha X\;.
\end{align}
The (tensorial) value of $w$ for a covector is $w=1-(\lambda,\lambda)=-{1\over4}$.

Define the Weizenb\"ock connection and its $Spin(5,5)$ and scaling components as
\begin{align}
\Gamma_{\alpha\beta}{}^\gamma=-(\*_\alpha EE^{-1})_\beta{}^\gamma
=-{1\over4}(\gamma_{ab})_\beta{}^\gamma\Gamma_\alpha{}^{ab}
+w\delta_\beta^\gamma\gamma_\alpha\;.
\end{align}
The Maurer--Cartan equations for $\Gamma$ read
\begin{align}
\*_{[\alpha}\Gamma_{\beta]}{}^{ab}&=-\Gamma_{[\alpha}{}^{c[a}\Gamma_{\beta]c}{}^{b]}\;,\nn\\
\*_{[\alpha}\gamma_{\beta]}&=0\;.
\end{align}

Torsion is defined as the tensorial part of the connection.
The inhomogeneous part of the transformation of the connection components under generalised diffeomorphisms is
\begin{align}
\Delta_\xi\Gamma_\alpha{}^{ab}&
     =-{1\over2}(\gamma^{ab})_\beta{}^\gamma\*_\alpha\*_\gamma\xi^\beta\;,\nn\\
\Delta_\xi\gamma_\alpha&=-\*_\alpha\*_\beta\xi^\beta\;.
\end{align}
It is then straightforward to verify that the torsion is given as
\begin{align}
\Theta_a{}^\alpha&=(\gamma^b\Gamma_{ab})^\alpha
   -{1\over10}(\gamma_a\gamma^{bc}\Gamma_{bc})^\alpha\;,\nn\\
\theta_\alpha&=-{1\over4}(\gamma^{ab}\Gamma_{ab})_\alpha-{5\over4}\gamma_\alpha\;,
\end{align}
where $\Theta$ is an irreducible vector-spinor.

The (generalised) torsion will obey Bianchi identities. The modules appearing can be read from level $-2$ of the tensor hierarchy algebra $W(E_5)$. We are in particular interested in the antisymmetric module, which 
will be the key to an action which is invariant under local rotations.
The Bianchi identities will be satisfied by the use of the Maurer--Cartan equation as well as the section constraint, which immediately states that $(\*\gamma_a\Gamma_{bc})=0$, $(\*\gamma_a\gamma)=0$.
Looking for a Bianchi identity in $(00100)$, one easily observes that the $\gamma^{(5)}$ terms cancel in
the combination
\begin{align}
(\*\gamma_{[ab}\Theta_{c]})-{2\over5}(\*\gamma_{abc}\theta)
=(\*\gamma_{[ab}{}^d\Gamma_{c]d})+{1\over2}(\*\gamma_{abc}\gamma)\;.
\end{align}
To make this into a covariant Bianchi identity, we need to add connection terms to the derivatives on the left hand side (with the representation matrices of eq. \eqref{RepMatrEq} and weight $-{1\over4}$), and use the Maurer--Cartan equation in the terms on the right hand side. Then all terms should collect into (torsion)${}^2$.
A direct but rather lengthy calculation leads to the result
\begin{align}
(D\gamma_{[ab}\Theta_{c]})-{2\over5}(D\gamma_{abc}\theta)
=-(\theta\gamma_{[ab}\Theta_{c]})\;.
\label{E5ASBI}
\end{align}
Note that there is no $\Theta^2$ term. Since $\vee^2(10010)$ does not contain $(00100)$, such terms are impossible.
The Bianchi identity can be written, 
with $X_{abc}{}^\alpha=(\gamma_{[ab}\Theta_{c]}-{2\over5}\gamma_{abc}\theta)^\alpha$, as
\begin{align}
(D_\alpha+\theta_\alpha)X_{abc}{}^\alpha=0\;.
\end{align}

Let $v^\alpha$ be a vector density of weight $(\lambda,\lambda)={5\over4}$. 
Then, $D_\alpha v^\alpha=(\*_\alpha-\theta_\alpha)v^\alpha$. The naked divergence is covariant. This means that if we (for purely formal reasons) introduce a covariantly constant tensor density $K^{abc}$ of weight $2(\lambda,\lambda)-1={3\over2}$, the Bianchi identity may further be rewritten as
\begin{align}
\*_\alpha(K^{abc}X_{abc}{}^\alpha)=0\;.
\end{align}
This kind of behaviour should be a generic phenomenon for the antisymmetric Bianchi identity. It will be useful for finding an action. The precise coefficient in the right hand side of eq. \eqref{E5ASBI} is crucial.

\subsection{Teleparallel dynamics}

Since torsion transforms covariantly under generalised diffeomorphisms, we only need to demand invariance under local $Spin(5)\times Spin(5)$ transformations.
Consider an infinitesimal $\mathfrak{so}(5)\oplus\mathfrak{so}(5)$ transformation of the vielbein, with parameter $\Lambda$, $\delta_\Lambda E=-E\Lambda$, leading to 
$\delta_\Lambda \Gamma_\alpha=E\*_\alpha\Lambda E^{-1}$. Converting
the flat indices to coordinate basis with $E$, so that we write 
$\Lambda_\alpha{}^\beta=(E\Lambda E^{-1})_\alpha{}^\beta$, we get, since the vielbein is covariantly constant, $\delta_\Lambda \Gamma_{\alpha\beta}{}^\gamma=D_\alpha\Lambda_\beta{}^\gamma$.
The characterisation of $\Lambda$ with coordinate indices is that it belongs to the locally defined compact subalgebra, it has eigenvalue $1$ under the involution defined by the metric, which is equivalently stated as $G^{-1}\Lambda$ being antisymmetric. We hope to find this antisymmetric parameter multiplying the Bianchi identity above after partial integration of the variation of the action.

In order to have a Lagrangian which is a scalar density of weight $1$, we let $E$ have 
weight ${1\over2}-(\lambda,\lambda)=-{3\over4}$. 
We search a Lagrangian $\LL$ which upon local $K$ transformations yields the Bianchi identity as
\begin{align}
\delta_\Lambda\LL\sim D_\alpha\Lambda^{abc}X_{abc}{}^\alpha
=\*_\alpha(\Lambda^{abc}X_{abc}{}^\alpha)\;,
\end{align}
where $\Lambda$ has been converted to a $3$-form with $K$-invariant tensors.

It may seem that in order to find the object which is linear in connections and that varies as $D\Lambda$, we need the full connection, and that the project would fail for the same reason that the spin connection is not fully determined by the torsion. On the other hand, $D\Lambda$ does not enter in arbitrary combinations, but contracts only torsion, so the modules corresponding to the undefined spin connection will not appear. Therefore, this is likely to succeed.

Let us try to write down a general Ansatz, and begin by counting possible terms.
Under the subgroup $Spin(5)\times Spin(5)\subset Spin(5,5)$, the torsion modules decompose as
\begin{align}
(10010)=\rep{144}&\rightarrow(\rep{16},\rep4)\oplus(\rep4,\rep{16})\oplus(\rep4,\rep4)\;,\nn\\
(00001)=\overline{\rep{16}}&\rightarrow(\rep4,\rep4)\;.
\end{align}
Therefore, there are three $K$-invariant terms $\Theta^2$, and one each of $\theta\Theta$ and $\theta^2$.
Their $G$-covariant forms are found by examining which modules in the symmetric products of torsion modules contain a $K$ singlet. They are
\begin{align}
\vee^2(10010)&\supset(00002)\oplus(00020)\oplus(20020)\;,\nn\\
(10010)\otimes(00001)&\supset(00020)\;,\\
\vee^2(00001)&\supset(00002)\;.\nn
\end{align}
A $(00002)$ can be contracted by $(G^{-1})^{abcde}$ to form an invariant. A $(00020)$ is contracted
with $G^{abcde}$, and a $(20020)$ by $(H^{-1})^{ab}G^{cdefg}$ (which of course also picks up $(00002)$).
Curiously, the terms with $G^{-1}$ and $H^{-1}G$ give the same weight, while the ones with $G$ gives another one (unless the weight of $G$ is $0$), and it is difficult to see how they can communicate in an action. The three terms turn out to be enough.

Consider the three terms in an Ansatz for the Lagrangian:
\begin{align}
A&=G^{abcde}(H^{-1})^{fg}(\Theta_f\gamma_{abcde}\Theta_g)\;,\nn\\
B&=(G^{-1})^{abcde}(\Theta_a\gamma_{bcd}\Theta_e)\;,\\
C&=(G^{-1})^{abcde}(\theta\gamma_{abcde}\theta)\;.\nn
\end{align}
In order for the Lagrangian to have the right weight, the weight of
the vielbein $E$ should be 
${1\over2}-(\lambda,\lambda)=-{3\over4}$.
The local transformation is parametrised by an (antisymmetric) tensor $\Lambda_{ab}$. It lies in $\mathfrak k$, and this is implemented by the condition that $H_{(a}{}^c\Lambda_{b)c}=0$.
With suitable normalisation, the connection transforms as
$\delta_\Lambda\Gamma_{\alpha ab}=D_\alpha\Lambda_{ab}$ (and, of course, 
$\delta_\Lambda\gamma_\alpha=0$), leading to
\begin{align}
\delta_\Lambda\Theta_a{}^\alpha&=(\gamma^bD)^\alpha\Lambda_{ab}
            -{1\over10}(\gamma_a\gamma^{bc}D)^\alpha\Lambda_{bc}\;,\nn\\
\delta_\Lambda\theta_\alpha&=-{1\over4}(\gamma^{ab}D)_\alpha\Lambda_{ab}\;.
\end{align}

We now calculate the variation of each of the three terms.
\begin{align}
\delta_\Lambda A
&=2G^{abcde}(H^{-1})^{fg}(D\Lambda_{fh}\gamma^h\gamma_{abcde}\Theta_g)
+{1\over5}G^{abcde}(H^{-1})^{fg}(D\Lambda_{hi}\gamma^{hi}\gamma_f\gamma_{abcde}\Theta_g)\nn\\
&=20G^{abcde}(H^{-1})^{fg}(D\Lambda_{fa}\gamma_{bcde}\Theta_g)
+2G^{abcde}(H^{-1})_a{}^g(D\Lambda_{hi}\gamma^{hi}\gamma_{bcde}\Theta_g)\nn\\
&=-20(G^{-1})^{abcde}(D\Lambda_a{}^f\gamma_{bcde}\Theta_f)
+2(G^{-1})^{abcde}(D\Lambda_{fg}\gamma^{fg}\gamma_{abcd}\Theta_e)\\
&=-20(G^{-1})^{abcde}(D\Lambda_a{}^f\gamma_{bcde}\Theta_f)\;.\nn
\end{align}
In the first step, a $\gamma$ matrix is taken through $\gamma^{(5)}$ to give contractions and a term that vanishes due to the wrong chirality, $G^{abcde}(\gamma_{abcde}\gamma_f\Theta_g)^\alpha=0$
(any such representation-theoretic identity can of course be proven concretely by double dualisation of the metric and the $\gamma$ matrices).
In the second step, $GH^{-1}$ is turned into $G^{-1}$ due to a contraction of indices, in the first term after using the antisymmetry of $(H^{-1}\Lambda)$.
Finally, the second term in the third line vanishes thanks to 
\begin{align}
(G^{-1})^{abcde}(\gamma_{abcd}\Theta_e)^\alpha
={1\over10}(G^{-1})^{abcde}(\gamma^f\gamma_{abcde}\Theta_f)^\alpha=0\;.
\end{align}
The remaining terms vary as
\begin{align}
\delta_\Lambda B
&=2(G^{-1})^{abcde}(D\Lambda_{af}\gamma^f\gamma_{bcd}\Theta_e)
+{1\over5}(G^{-1})^{abcde}(D\Lambda_{fg}\gamma^{fg}\gamma_{abcd}\Theta_e)\nn\\
&=2(G^{-1})^{abcde}(D\Lambda_a{}^f\gamma_{fbcd}\Theta_e)
+6(G^{-1})^{abcde}(D\Lambda_{ab}\gamma_{cd}\Theta_e)\;,
\end{align}
and
\begin{align}
\delta_\Lambda C
&={1\over2}(G^{-1})^{abcde}(D\Lambda_{fg}\gamma^{fg}\gamma_{abcde}\theta)\nn\\
&=-5(G^{-1})^{abcde}(D\Lambda_a{}^f\gamma_f\gamma_{bcde}\theta)\nn\\
&=-5(G^{-1})^{abcde}(D\Lambda_a{}^f\gamma_{fbcde}\theta)
-20(G^{-1})^{abcde}(D\Lambda_{ab}\gamma_{cde}\theta)\\
&=-20(G^{-1})^{abcde}(D\Lambda_{ab}\gamma_{cde}\theta)\;.\nn
\end{align}
The first term in the third line can be shown to vanish by inserting $H$ and $H^{-1}$ by\linebreak 
$\Lambda=-(H^{-1}\Lambda H)^t$, turning $HG^{-1}$ into $G$ and using
$G_{[a}{}^{bcde}(\gamma_{f]bcde}\theta)^\alpha=0$, since the product $(00020)\otimes(00020)$ does not contain $(01000)$.

If we now choose (up to an overall factor)
\begin{align}
\LL=-{1\over240}A+{1\over6}B+{1\over50}C\;,
\end{align}
the variation adds up to
\begin{align}
\delta_\Lambda\LL
&={1\over12}(G^{-1})^{abcde}(D\Lambda_a{}^f\gamma_{fbcd}\Theta_e)
+{1\over3}(G^{-1})^{abcde}(D\Lambda_a{}^f\gamma_{bcde}\Theta_f)\nn\\
&\quad+(G^{-1})^{abcde}(D\Lambda_{ab}\gamma_{cd}\Theta_e)
-{2\over5}(G^{-1})^{abcde}(D\Lambda_{ab}\gamma_{cde}\theta)\;.
\end{align}
The terms in the first row cancel, since
\begin{align}
(G^{-1})_a{}^{bcde}(\gamma_{[bcde}\Theta_{f]})^\alpha
={1\over10}(G^{-1})_{a}{}^{bcde}(\gamma^g\gamma_{bcdef}\Theta_g)^\alpha\;,
\end{align} 
which is symmetric in $(af)$.
Thus, the variation becomes a total derivative thanks to the Bianchi identity, and
\begin{align}
\delta_\Lambda\LL=\*_\alpha((G^{-1})^{abcde}\Lambda_{ab}X_{cde}{}^\alpha)\;.
\end{align}

We have not explicitly checked the equivalence to the usual (``coset'') formulation of $E_5$ exceptional field theory, since it obviously follows from the uniqueness of the respective Lagrangians.
If the generalised vielbein is parametrised in terms of ordinary
vielbein and $3$-form potential, the Lagrangian will give the
teleparallel version of gravity together with the dynamics of the
$3$-form.

\subsection{Local symmetry and irreducibility as dual gauge
  symmetries}


All fields, fields strengths, torsion components etc. come in modules
of the structure algebra $E_5$. Nevertheless, we have seen that a
specific Bianchi identity is responsible for the invariance under the
local subalgebra. The reason is that, with an appropriate definition
of the dualisation map of the field strengths (torsion) from the complex $C$ of Figure \ref{EFiveComplex} 
to the
dual complex $C^\star$, the Bianchi identity maps to a local symmetry for the
dual field strength. We call this a dual gauge symmetry. Note that this
does {\it not} involve a change of local degrees of freedom (the generalised
vielbein) to some dual variables, but relies on the transformation
being realisable as a local transformation of them.

The critical property of the anti-symmetric Bianchi identity is that
it takes the precise form 
\begin{align}
(D+\theta)_MX_{\scr A}{}^M\;,
\label{XBIForm}
\end{align}
 where ${\scr
  A}$ is an index for the module of the Bianchi identity. Then,
a variation of an action yielding $D_M\Lambda^{\scr A}X_{\scr A}{}^M$
gives a total derivative.

The leading antisymmetric module in the Bianchi identity is always
accompanied by a subleading symmetric module. There are in fact two
such Bianchi identities, one which vanishes on lowering (the adjoint
action of $e_{-1}$) an one that
vanishes on raising (the opposite operation), like the antisymmetric
one. We are not clear on
the precise significance of this property. In the situations where the
tensor hierarchy algebra possesses a non-degenerate quadratic form, it
implies that these modules do not appear in the (would-be) dual
position in $W(\fg)$. 

Does the subleading symmetric module have the
same property \eqref{XBIForm}, so that it may correspond to a total derivative in an
action?
We have checked this for $\fg=E_5$. With the definitions given above
of the torsion components, one may derive the Bianchi identities in
the subleading symmetric module, the $so(10)$ vector $\bf10$.
They become
\begin{align}
  (D\Theta_a)&=-\fr{10}(\Theta_b\gamma_a\Theta^b)
  -\Fr{31}{25}(\theta\Theta_a)+\Fr{36}{125}(\theta\gamma_a\theta)\;,\nn\\
  (D\gamma_a\theta)&=-\fr4(\Theta_b\gamma_a\Theta^b)
  -\Fr35(\theta\Theta_a)+\Fr7{25}(\theta\gamma_a\theta)\;.
\end{align}
One linear combination of these identities, the one without $\Theta^2$, is
\begin{align}
(D+\theta)(\Theta_a-\Fr25\gamma_a\theta)=0\;,
\end{align}

Suppose that the vielbein is subject to a variation
$\delta_\Sigma E=-E\Sigma$, where $\Sigma$ is an element in the ``coset
directions'' $\fg\ominus\fk$ such that
$(G^{-1}t_\alpha)^{[MN]}\Sigma^\alpha=
(G^{-1}t_\alpha)^{\langle MN\rangle}\Sigma^\alpha=0$.
Then, the transformation with parameter $\Sigma$ would serve to remove the
subleading symmetric part of the metric. In the analysis above, we
have already assumed that the metric is in the leading symmetric
module. The present observation indicates that the metric could be
taken as a general symmetric matrix, and that the unphysical
subleading parts can be gauged away using such a symmetry.
Note that this phenomenon is absent in ordinary gravity.

\section{Dynamics for the general situation with irreducible torsion\label{GeneralDynamics}}

Inspired by the $Spin(5,5)$ formulation of teleparallel dynamics, we will now give a general form of the action, for simplicity in the situation when the big torsion module is irreducible, \ie, when $\varphi$ projects on a single irreducible module. Note that this implies that $\ell=0$ --- equivalently, $\lambda$ is dual to a simple root 
$\alpha$ with Coxeter label $1$ \cite{Cederwall:2017fjm} --- but also that $\alpha$ is an ``outer'' root, so that the Dynkin diagram with the node $\lambda$ connects to removed is connected. This includes essentially all models of interest without ancillary transformations.

The idea is that the big torsion module will enter through two types of terms in the action, 
$\eta_{\alpha\beta}G^{MN}\Theta_M{}^\alpha\Theta_N{}^\beta$ and
$G_{\alpha\beta}G^{MN}\Theta_M{}^\alpha\Theta_N{}^\beta$. Here, $G_{\alpha\beta}$ is the metric on the adjoint module, and can (with one index raised, which we always do with $\eta$) be thought of as (minus) the actual involution on the algebra. The defining relation is
\begin{align}
G^{\alpha\beta}G_{MP}G^{NQ}t_{\beta Q}{}^P=t^\alpha{}_M{}^N\;.
\label{InvolutionDef}
\end{align}
Note that this metric is its own inverse, $G_\alpha{}^\gamma G_\gamma{}^\beta=\delta_\alpha^\beta$, and that it carries weight $0$, independently of the weight of $G_{MN}$.
The purpose of the two terms is to form singlets together with the square of $\Theta$ in the modules
$R(2\lambda)$ and $R(2(\lambda+\gamma))\subset R(2\lambda)\otimes R(2\theta)$ (here, $\gamma$ is the unique highest root of $\fg$ such that $(\lambda,\gamma)=0$), which will both contain singlets under the compact subgroup $K$.

In the absence of ancillary transformations, $\ell=0$, and the invariant tensor $\varphi$ takes the form
\begin{align}
\varphi^\alpha{}_\beta&=\delta^\alpha_\beta+f^\alpha{}_\beta{}^\gamma t_\gamma
-\fr{(\lambda,\lambda)}t^\alpha t_\beta\nn\\
&=\delta^\alpha_\beta-t_\beta t^\alpha
+(1-\fr{(\lambda,\lambda)})t^\alpha t_\beta\;.
\end{align}
Note that the last term (in both lines) vanishes when acting on $\Theta^\beta$, since
$t_\alpha\Theta^\alpha=0$.
Another important property is $G_{\alpha\beta}t^\alpha\Theta^\beta=0$. This identity is derived by using eq. \eqref{InvolutionDef} and using the fact that $G^{MN}$ is in the leading module $R(-2\lambda)$. Then, the indices on $G^{-1}$ contract lower indices on $t$ and $\Theta$. This vanishes due to the conjugate of eq. \eqref{tPhiZero}.
The eigenvalue of $\varphi$ acting on $\Theta$ in the (single) module $R(\lambda+\gamma)$, $\gamma$ being the highest root with $(\lambda,\gamma)=0$, was calculated in ref. 
\cite{Cederwall:2019qnw}; it is $N=g^\vee-1-(\gamma,\varrho)$.

The variation of the connection under a local $\mathfrak k$ transformation is 
$\delta_\Lambda\Gamma_M{}^\alpha=D_M\Lambda^\alpha$, leading to
$\delta_\Lambda\Theta_M{}^\alpha=\varphi^\alpha{}_{M,\beta}{}^ND_N\Lambda^\beta$.
The fact that $\Lambda\in{\mathfrak k}$ is expressed by 
$G^\alpha{}_\beta\Lambda^\beta=-\Lambda^\alpha$, or equivalently
$G^{P(M}t_{\alpha P}{}^{N)}\Lambda^\alpha=0$. We write
$\Lambda^{MN}=G^{MP}t_{\alpha P}{}^{N}\Lambda^\alpha$.

We can now check the transformations of the two proposed terms.
A short calculation, using the explicit form of $\varphi$ together with $\Lambda\in\mathfrak k$ and eq. \eqref{InvolutionDef}, leads to (in notation with fundamental indices suppressed)
\begin{align}
\delta_\Lambda(\fr2G^{-1}\eta_{\alpha\beta}\Theta^\alpha\otimes\Theta^\beta)
&=G^{-1}(\varphi_{\beta\alpha}\otimes1)D\Lambda^\alpha\otimes\Theta^\beta\nn\\
&=G^{-1}\eta_{\alpha\beta}D\Lambda^\alpha\otimes\Theta^\beta
+G^{-1}(t_\beta\otimes t_\alpha)D\Lambda^\alpha\otimes\Theta^\beta\;,\nn\\
\delta_\Lambda(\fr2G^{-1}G_{\alpha\beta}\Theta^\alpha\otimes\Theta^\beta)
&=G^{-1}G_{\alpha\beta}(\varphi^\alpha{}_\gamma\otimes1)D\Lambda^\gamma\otimes\Theta^\beta\\
&=-G^{-1}(1\otimes\varphi_{\alpha\beta})D\Lambda^\alpha\otimes\Theta^\beta\nn\\
&=-NG^{-1}\eta_{\alpha\beta}D\Lambda^\alpha\otimes\Theta^\beta\;.\nn
\end{align}
The correct term with $\Theta$ in the Bianchi identity is obtained by partially integrating
\begin{align}
\delta_\Lambda(\fr2G^{-1}(\eta_{\alpha\beta}+\fr NG_{\alpha\beta})
\Theta^\alpha\otimes\Theta^\beta)
=G^{-1}(t_\beta\otimes t_\alpha)D\Lambda^\alpha\otimes\Theta^\beta\;.
\end{align}
To this is added the transformation of $G^{-1}\theta\otimes\theta$, 
\begin{align}
\delta_\Lambda(\fr2G^{-1}\theta\otimes\theta)=G^{-1}(t_\alpha\otimes1)D\Lambda^\alpha\otimes\theta\;.
\end{align}
Using the Bianchi identity \eqref{LeadingBI}, we obtain
\begin{align}
&\delta_\Lambda(\fr2G^{-1}(\eta_{\alpha\beta}+\fr NG_{\alpha\beta})\Theta^\alpha\otimes\Theta^\beta
+(2-\fr{(\lambda,\lambda)})G^{-1}\theta\otimes\theta)\nn\\
&=\*_P\bigl(\Lambda^{MN}(t_{\alpha M}{}^P\Theta_N{}^\alpha
+2(2-\fr{(\lambda,\lambda)})\delta^P_M\theta_N)\bigr)\;.
\end{align}
This holds provided that $\Lambda^{MN}$ does not contain any subleading antisymmetric module.
 For $\fg=E_7$, $R(\lambda)={\bf56}$, there is a subleading antisymmetric module, the singlet, but it is 
of course not contained in $R(-2\lambda)\otimes\hbox{\bf adj}$ (the factors representing $G^{-1}$ and $\Lambda$).

\section{Conclusions and outlook\label{Conclusions}}

We have demonstrated how to give a teleparallel formulation of
extended geometry. The crucial observation is the generic existence of
an antisymmetric torsion Bianchi identity, and that the precise
non-linear form of this Bianchi identity is such that it matches the
form of a covariant divergence. This ensures the symmetry of an action
under local rotations in the compact subgroup. 
We have also found indications, not further elaborated on, that any
subleading symmetric part of a metric can be gauged away in a similar
fashion. 
It is striking that these Bianchi identities, the ones with the
potential to remove local degrees of freedom from the generalised
vielbein, are found precisely as the parts of $S(\fg^+)$ at level
$(p,q)=(-2,0)$ which can be lowered to $(p,q)=(-2,-1)$. This is probably
significant, and there may be a reason to expect the possibility of a
general proof of the forms of these Bianchi identities in terms of the
brackets of the tensor hierarchy algebra.  

The formalism obtained is ideally suited to the view that extended
geometry is constructed from a tensor hierarchy algebra.   
A natural continuation of this work is to extend it to
infinite-dimensional structure groups, in particular over-extended
ones.  
On the way there, one passes the cases of adjoint and affine structure
groups. We have partial encouraging results for these. In the adjoint
(Ehlers) situation, the ``big'' torsion module is always reducible,
since it contains a singlet (in some cases, it is further
reducible). Also, the antisymmetric module is reducible. We have
verified that also the subleading antisymmetric module, the adjoint,
contains a Bianchi identity of the required form. 
In the affine (Geroch) case,
we have observed that the tensor hierarchy algebra predicts the existence
of an ancillary field, in agreement with ref. \cite{Bossard:2018utw}, and
that there is an antisymmetric Bianchi identity of the correct
form. This will be reported in a forthcoming paper. 

For infinite-dimensional structure groups, the present formalism should make it possible to view the field strengths (torsion) as arising from a group element, which is the exponentiation 
$G$ of an algebra element at 
$(p,q)=(0,0)$. Generically, this group is larger than the exponentiation of $\fg$, due to the presence of ``extra'' elements \cite{Cederwall:2021ymp}, starting with the Virasoro generator $L_1$ for affine $\fg$. Such an approach would remedy the difficulties, arising in \eg\ refs. \cite{Bossard:2017wxl,Bossard:2021jix,Bossard:2021ebg}, associated with assigning field to the coadjoint module of $G$ instead of the adjoint one, and allow for exposure of the full local symmetries of the theories.

\section*{Acknowledgements}

MC would like to thank Axel Kleinschmidt for discussions.


\bibliographystyle{utphysmod2}



\begin{thebibliography}{10}

\bibitem{Cederwall:2017fjm}
M.~Cederwall and J.~Palmkvist,  {\em {Extended geometries}}, JHEP {\bf 02}, 071
  (2018)
[\href{http://www.arXiv.org/abs/1711.07694}{{\tt 1711.07694}}].

\bibitem{Cederwall:2018aab}
M.~Cederwall and J.~Palmkvist,  {\em {$L_{\infty }$ algebras for extended
  geometry from Borcherds superalgebras}}, Commun. Math. Phys. {\bf 369},
  721--760 (2019)
[\href{http://www.arXiv.org/abs/1804.04377}{{\tt 1804.04377}}].

\bibitem{Cederwall:2019qnw}
M.~Cederwall and J.~Palmkvist,  {\em {Tensor hierarchy algebras and extended
  geometry. Part I. Construction of the algebra}}, JHEP {\bf 02}, 144 (2020)
  [\href{http://www.arXiv.org/abs/1908.08695}{{\tt 1908.08695}}].

\bibitem{Cederwall:2019bai}
M.~Cederwall and J.~Palmkvist,  {\em {Tensor hierarchy algebras and extended
  geometry. Part II. Gauge structure and dynamics}}, JHEP {\bf 02}, 145 (2020)
  [\href{http://www.arXiv.org/abs/1908.08696}{{\tt 1908.08696}}].

\bibitem{Hull:2007zu}
C.~M. Hull,  {\em Generalised geometry for {M}-theory}, JHEP {\bf 0707}, 079
  (2007)
[\href{http://www.arXiv.org/abs/hep-th/0701203}{{\tt hep-th/0701203}}].

\bibitem{Pacheco:2008ps}
P.~P. Pacheco and D.~Waldram,  {\em {M}-theory, exceptional generalised
  geometry and superpotentials}, JHEP {\bf 0809}, 123 (2008)
[\href{http://www.arXiv.org/abs/0804.1362}{{\tt 0804.1362}}].

\bibitem{Hillmann:2009pp}
C.~Hillmann,  {\em {$E_{7(7)}$ and $d=11$ supergravity}},
  \href{http://www.arXiv.org/abs/0902.1509}{{\tt 0902.1509}}.
PhD thesis, Humboldt-Universit\"at zu Berlin, 2008.

\bibitem{Berman:2010is}
D.~S. Berman and M.~J. Perry,  {\em Generalized geometry and {M} theory}, JHEP
  {\bf 1106}, 074 (2011)
[\href{http://www.arXiv.org/abs/1008.1763}{{\tt 1008.1763}}].

\bibitem{Berman:2011pe}
D.~S. Berman, H.~Godazgar and M.~J. Perry,  {\em {$SO(5,5)$} duality in
  {M}-theory and generalized geometry}, Phys.Lett. {\bf B700}, 65--67 (2011)
[\href{http://www.arXiv.org/abs/1103.5733}{{\tt 1103.5733}}].

\bibitem{Coimbra:2011ky}
A.~Coimbra, C.~Strickland-Constable and D.~Waldram,  {\em {$E_{d(d)} \times
  \mathbb{R}^+$ generalised geometry, connections and M theory}}, JHEP {\bf
  1402}, 054 (2014)
[\href{http://www.arXiv.org/abs/1112.3989}{{\tt 1112.3989}}].

\bibitem{Coimbra:2012af}
A.~Coimbra, C.~Strickland-Constable and D.~Waldram,  {\em Supergravity as
  generalised geometry {II}: {$E_{d(d)} \times \mathbb{R}^+$} and {M} theory},
  JHEP {\bf 1403}, 019 (2014)
[\href{http://www.arXiv.org/abs/1212.1586}{{\tt 1212.1586}}].

\bibitem{Berman:2012vc}
D.~S. Berman, M.~Cederwall, A.~Kleinschmidt and D.~C. Thompson,  {\em {The
  gauge structure of generalised diffeomorphisms}}, JHEP {\bf 01}, 064 (2013)
[\href{http://www.arXiv.org/abs/1208.5884}{{\tt 1208.5884}}].

\bibitem{Park:2013gaj}
J.-H. Park and Y.~Suh,  {\em {U}-geometry : {SL(5)}}, JHEP {\bf 04}, 147 (2013)
[\href{http://www.arXiv.org/abs/1302.1652}{{\tt 1302.1652}}].

\bibitem{Cederwall:2013naa}
M.~Cederwall, J.~Edlund and A.~Karlsson,  {\em {Exceptional geometry and tensor
  fields}}, JHEP {\bf 07}, 028 (2013)
[\href{http://www.arXiv.org/abs/1302.6736}{{\tt 1302.6736}}].

\bibitem{Cederwall:2013oaa}
M.~Cederwall,  {\em {Non-gravitational exceptional supermultiplets}}, JHEP {\bf
  07}, 025 (2013)
[\href{http://www.arXiv.org/abs/1302.6737}{{\tt 1302.6737}}].

\bibitem{Aldazabal:2013mya}
G.~Aldazabal, M.~Gra\~{n}a, D.~Marqu\'es and J.~A. Rosabal,  {\em {Extended
  geometry and gauged maximal supergravity}}, JHEP {\bf 1306}, 046 (2013)
[\href{http://www.arXiv.org/abs/1302.5419}{{\tt 1302.5419}}].

\bibitem{Hohm:2013pua}
O.~Hohm and H.~Samtleben,  {\em Exceptional form of ${D}=11$ supergravity},
  Phys. Rev. Lett. {\bf 111}, 231601 (2013)
[\href{http://www.arXiv.org/abs/1308.1673}{{\tt 1308.1673}}].

\bibitem{Blair:2013gqa}
C.~D. Blair, E.~Malek and J.-H. Park,  {\em {M}-theory and type {IIB} from a
  duality manifest action}, JHEP {\bf 1401}, 172 (2014)
[\href{http://www.arXiv.org/abs/1311.5109}{{\tt 1311.5109}}].

\bibitem{Hohm:2013vpa}
O.~Hohm and H.~Samtleben,  {\em Exceptional field theory {I}: {E}$_{6(6)}$
  covariant form of {M}-theory and type {IIB}}, Phys. Rev. {\bf D89}, 066016
  (2014)
[\href{http://www.arXiv.org/abs/1312.0614}{{\tt 1312.0614}}].

\bibitem{Hohm:2013uia}
O.~Hohm and H.~Samtleben,  {\em Exceptional field theory {II}: {E}$_{7(7)}$},
  Phys. Rev. {\bf D89}, 066017 (2014)
[\href{http://www.arXiv.org/abs/1312.4542}{{\tt 1312.4542}}].

\bibitem{Hohm:2014fxa}
O.~Hohm and H.~Samtleben,  {\em {Exceptional field theory. III. E$_{8(8)}$}},
  Phys. Rev. {\bf D90}, 066002 (2014)
[\href{http://www.arXiv.org/abs/1406.3348}{{\tt 1406.3348}}].

\bibitem{Cederwall:2015ica}
M.~Cederwall and J.~A. Rosabal,  {\em {E$_{8}$ geometry}}, JHEP {\bf 07}, 007
  (2015)
[\href{http://www.arXiv.org/abs/1504.04843}{{\tt 1504.04843}}].

\bibitem{Butter:2018bkl}
D.~Butter, H.~Samtleben and E.~Sezgin,  {\em {E$_{7(7)}$ exceptional field
  theory in superspace}}, JHEP {\bf 01}, 087 (2019)
[\href{http://www.arXiv.org/abs/1811.00038}{{\tt 1811.00038}}].

\bibitem{Bossard:2017aae}
G.~Bossard, M.~Cederwall, A.~Kleinschmidt, J.~Palmkvist and H.~Samtleben,  {\em
  {Generalized diffeomorphisms for $E_9$}}, Phys. Rev. {\bf D96}, 106022 (2017)
[\href{http://www.arXiv.org/abs/1708.08936}{{\tt 1708.08936}}].

\bibitem{Bossard:2018utw}
G.~Bossard, F.~Ciceri, G.~Inverso, A.~Kleinschmidt and H.~Samtleben,  {\em
  {E$_9$ exceptional field theory I. The potential}},
\href{http://www.arXiv.org/abs/1811.04088}{{\tt 1811.04088}}.

\bibitem{Bossard:2019ksx}
G.~Bossard, A.~Kleinschmidt and E.~Sezgin,  {\em {On supersymmetric E$_{11}$
  exceptional field theory}},
\href{http://www.arXiv.org/abs/1907.02080}{{\tt 1907.02080}}.

\bibitem{Bossard:2021jix}
G.~Bossard, F.~Ciceri, G.~Inverso, A.~Kleinschmidt and H.~Samtleben,  {\em
  {E$_{9}$ exceptional field theory. Part II. The complete dynamics}}, JHEP
  {\bf 05}, 107 (2021) [\href{http://www.arXiv.org/abs/2103.12118}{{\tt
  2103.12118}}].

\bibitem{Bossard:2021ebg}
G.~Bossard, A.~Kleinschmidt and E.~Sezgin,  {\em {A master exceptional field
  theory}}, JHEP {\bf 06}, 185 (2021)
  [\href{http://www.arXiv.org/abs/2103.13411}{{\tt 2103.13411}}].

\bibitem{Tseytlin:1990va}
A.~A. Tseytlin,  {\em {Duality symmetric closed string theory and interacting
  chiral scalars}}, Nucl. Phys. {\bf B350}, 395--440
(1991).

\bibitem{Siegel:1993xq}
W.~Siegel,  {\em {Two vierbein formalism for string inspired axionic gravity}},
  Phys. Rev. {\bf D47}, 5453--5459 (1993)
[\href{http://www.arXiv.org/abs/hep-th/9302036}{{\tt hep-th/9302036}}].

\bibitem{Siegel:1993bj}
W.~Siegel,  {\em {Manifest duality in low-energy superstrings}}, in {\em
  {International Conference on Strings 93 Berkeley, California, May 24-29,
  1993}}, pp.~353--363.
\newblock 1993.
\newblock
\href{http://www.arXiv.org/abs/hep-th/9308133}{{\tt hep-th/9308133}}.
\newblock

\bibitem{Hitchin:2010qz}
N.~Hitchin,  {\em {Lectures on generalized geometry}},
\href{http://www.arXiv.org/abs/1008.0973}{{\tt 1008.0973}}.

\bibitem{Hull:2004in}
C.~M. Hull,  {\em {A geometry for non-geometric string backgrounds}}, JHEP {\bf
  10}, 065 (2005)
[\href{http://www.arXiv.org/abs/hep-th/0406102}{{\tt hep-th/0406102}}].

\bibitem{Hull:2006va}
C.~M. Hull,  {\em {Doubled geometry and T-folds}}, JHEP {\bf 07}, 080 (2007)
[\href{http://www.arXiv.org/abs/hep-th/0605149}{{\tt hep-th/0605149}}].

\bibitem{Hull:2009mi}
C.~M. Hull and B.~Zwiebach,  {\em {Double field theory}}, JHEP {\bf 09}, 099
  (2009)
[\href{http://www.arXiv.org/abs/0904.4664}{{\tt 0904.4664}}].

\bibitem{Hohm:2010jy}
O.~Hohm, C.~M. Hull and B.~Zwiebach,  {\em {Background independent action for
  double field theory}}, JHEP {\bf 07}, 016 (2010)
[\href{http://www.arXiv.org/abs/1003.5027}{{\tt 1003.5027}}].

\bibitem{Hohm:2010pp}
O.~Hohm, C.~M. Hull and B.~Zwiebach,  {\em {Generalized metric formulation of
  double field theory}}, JHEP {\bf 08}, 008 (2010)
[\href{http://www.arXiv.org/abs/1006.4823}{{\tt 1006.4823}}].

\bibitem{Jeon:2012hp}
I.~Jeon, K.~Lee, J.-H. Park and Y.~Suh,  {\em {Stringy unification of type IIA
  and IIB supergravities under $N=2$ $D=10$ supersymmetric double field
  theory}}, Phys. Lett. {\bf B723}, 245--250 (2013)
[\href{http://www.arXiv.org/abs/1210.5078}{{\tt 1210.5078}}].

\bibitem{Park:2013mpa}
J.-H. Park,  {\em {Comments on double field theory and diffeomorphisms}}, JHEP
  {\bf 06}, 098 (2013)
[\href{http://www.arXiv.org/abs/1304.5946}{{\tt 1304.5946}}].

\bibitem{Berman:2014jba}
D.~S. Berman, M.~Cederwall and M.~J. Perry,  {\em {Global aspects of double
  geometry}}, JHEP {\bf 09}, 066 (2014)
[\href{http://www.arXiv.org/abs/1401.1311}{{\tt 1401.1311}}].

\bibitem{Cederwall:2014kxa}
M.~Cederwall,  {\em {The geometry behind double geometry}}, JHEP {\bf 09}, 070
  (2014)
[\href{http://www.arXiv.org/abs/1402.2513}{{\tt 1402.2513}}].

\bibitem{Cederwall:2014opa}
M.~Cederwall,  {\em {T-duality and non-geometric solutions from double
  geometry}}, Fortsch. Phys. {\bf 62}, 942--949 (2014)
[\href{http://www.arXiv.org/abs/1409.4463}{{\tt 1409.4463}}].

\bibitem{Cederwall:2016ukd}
M.~Cederwall,  {\em {Double supergeometry}}, JHEP {\bf 06}, 155 (2016)
[\href{http://www.arXiv.org/abs/1603.04684}{{\tt 1603.04684}}].

\bibitem{Ehlers:1957zz}
J.~Ehlers,  {\em {Konstruktionen und Charakterisierung von L\"osungen der
  Einsteinschen Gravitationsfeldgleichungen}},. PhD thesis, Hamburg, 1957.

\bibitem{Geroch:1972yt}
R.~P. Geroch,  {\em {A Method for generating new solutions of Einstein's
  equation. 2}}, J. Math. Phys. {\bf 13}, 394--404 (1972).

\bibitem{BKL}
V.~A. Belinskii, I.~M. Khalatnikov and E.~M. Lifshitz,  {\em Oscillatory
  approach to a singular point in the relativistic cosmology}, Adv. Phys. {\bf
  19}, 525 (1970).

\bibitem{Palmkvist:2013vya}
J.~Palmkvist,  {\em {The tensor hierarchy algebra}}, J. Math. Phys. {\bf 55},
  011701 (2014)
[\href{http://www.arXiv.org/abs/1305.0018}{{\tt 1305.0018}}].

\bibitem{Palmkvist:2015dea}
J.~Palmkvist,  {\em {Exceptional geometry and Borcherds superalgebras}}, JHEP
  {\bf 11}, 032 (2015)
[\href{http://www.arXiv.org/abs/1507.08828}{{\tt 1507.08828}}].

\bibitem{Carbone:2018xqq}
L.~Carbone, M.~Cederwall and J.~Palmkvist,  {\em {Generators and relations for
  Lie superalgebras of Cartan type}}, J. Phys. {\bf A52}, 055203 (2019)
[\href{http://www.arXiv.org/abs/1802.05767}{{\tt 1802.05767}}].

\bibitem{Cederwall:2021ymp}
M.~Cederwall and J.~Palmkvist,  {\em {Tensor hierarchy algebra extensions of
  over-extended Kac--Moody algebras}}, {Commun. Math. Phys.} ({2021})
  [\href{http://www.arXiv.org/abs/2103.02476}{{\tt 2103.02476}}].

\bibitem{Einstein:1926}
A.~Einstein,  {\em {Riemann-Geometrie mit Aufrechterhaltung des Begriffes des
  Fernparallelismus}}, {Preussische Akademie der Wissenschaften, Phys.-math.
  Klasse, Sitzungsberichte} 217 (1926).

\bibitem{DeAndrade:2000sf}
V.~C. De~Andrade, L.~C.~T. Guillen and J.~G. Pereira,  {\em {Teleparallel
  gravity: An overview}}, in {\em {9th Marcel Grossmann meeting (MG 9)}}.
\newblock 11, 2000.
\newblock \href{http://www.arXiv.org/abs/gr-qc/0011087}{{\tt gr-qc/0011087}}.

\bibitem{Bossard:2017wxl}
G.~Bossard, A.~Kleinschmidt, J.~Palmkvist, C.~N. Pope and E.~Sezgin,  {\em
  {Beyond $E_{11}$}}, JHEP {\bf 05}, 020 (2017)
[\href{http://www.arXiv.org/abs/1703.01305}{{\tt 1703.01305}}].

\end{thebibliography}

\providecommand{\href}[2]{#2}\begingroup\raggedright\endgroup

\end{document}